\documentclass[1p]{elsarticle}

\usepackage{graphicx}
\usepackage{caption}
\usepackage{subcaption}
\usepackage{xcolor}
\usepackage[fleqn]{amsmath}
\usepackage{hyperref}
\usepackage{float}
\usepackage{dblfloatfix}
\usepackage{epstopdf}

\usepackage{setspace}
\newcommand*{\figref}[1]{Fig.~\ref{#1}}
\graphicspath{ {./EPS/} }

\begin{document}
	
\begin{frontmatter}
\title{Detection of true Gaussian shaped pulses at high count rates}
\author[add1]{M.Yu.~Kantor \corref{cor1}}
\ead{m.kantor@mail.ioffe.ru}

\author[add1]{A.V.~Sidorov}
\ead{sidorov@mail.ioffe.ru}	 

\address[add1]{Ioffe Institute, Polytekhnicheskaya 26, 194021, Saint-Petersburg, Russia}

\cortext[cor1]{Corresponding author}

\date{}

\begin{abstract}

A new true Gaussian digital shaper of detector pulses is tested and compared to the standard trapezoidal shaper in terms of count rate, amplitude resolution and biasing the output shaped pulses. The true Gaussian shaper allows for shaping detector pulses into a symmetrical form which width can be considerably less than the rise time of the input pulses. Therefore, the dead time of the true Gaussian shaper can be noticeably reduced in regards to that of trapezoidal shaper. Less dead time provides a higher count rate of the true Gaussian shapers. The true Gaussian and trapezoidal shapers are compared in a wide range of the input count rates. The maximal output count rate of the true Gaussian shaper has been found to exceed the rate of the trapezoidal shapers in several times. The true Gaussian shaper provides biasing-free measurements of pulse amplitudes  at high count rates with better resolution than that provided by the trapezoidal shapers. The tests have been performed by modeling the output signals of silicon drift detector (SDD) H7 VITUS equipped with AXAS-D pre-amplifier developed by KETEK GmbH for measurements of soft X-ray spectra at high count rates.  

\end{abstract}

\begin{keyword}
	Digital filters\sep pulse shaping\sep high count rate measurements\sep silicon drift detectors
\end{keyword}

\end{frontmatter}

\section{Introduction}
\label{sec:Introduction}

The impulse response of radiation detectors formed by a pre-amplifier has usually short rise time $\tau_r$ and long falling tail $\tau_t\gg\tau_r$ which durations are a trade-off between output noise, ballistic deficit and pulse pile-up of the detector system \cite{Knoll2010}. At high input rates, the response pulses are overlapped and badly resolved resulting in pile-up signals. The number of pile-up pulses is significantly reduced by shaping the impulse response into a symmetrical form of a shorter width $T_s<\tau_r+\tau_t$. The width of pulses is defined hereafter as the full width at its half maximum (FWHM). At $\tau_t>T_s\gg\tau_r$, standard shapers \cite{jordanov1994digital,jordanov1994digital1} provide well-formed symmetrical output pulses at a low noise level. The inferred amplitudes of the shaped pulses are unbiased at count rates lower than $\approx1/T_s$. At shorter pulse widths $T_s\leq 2\tau_r$, the output of standard shapers is disturbed from the symmetrical form, output noises are increased and the inferred pulse amplitudes are biased from their real values. A true Gaussian filter was recently developed to shorten the shaped pulses to less than the rise time of the response pulse \cite{Kantor2019,Kantor2018,Kantor2019Arx} in order to provide the highest count rates of pulse measurements. 

The output count rate, energy resolution and biasing amplitudes of the shapers are simulated in the paper for true Gaussian and trapezoidal pulses shaped from the impulse response $S_R$ of an AXAS-D spectrometer \cite{KETEKAXAS} of soft X-ray (SXR) radiation equipped with a silicon drift detector (SDD) H7 VITUS \cite{KETEKVITUS} of obsolete standard class. The spectrometer was developed by KETEK GmbH in 2012 for measurements of soft X-ray spectra at input count rates up to 5$\cdot 10^5$ 1/s. The best energy resolution of the spectrometer is 140 eV and it is provided and low input count rates and the peaking time of the response pulses 8 $mu s$. The pre-amplifier of the employed spectrometer delivers the response pulses with the peaking time 0.2$mu s$. This time is somewhat larger the rise time of the  step-like detector signals. This short peaking time allows a higher count rates of pulse detection but with resolution degraded to $>200$ eV. The impulse response was accurately measured \cite{Kantor2019} and used throughout the analysis in this paper. Note, modern silicon drift detectors of KETEK and AMPTEK exhibit less noise and several times higher count rates \cite{KETEKVITUSCube, AMPTEKSDD}. The application of the true Gaussian shaper in SXR spectrometers built with the modern SDD can be evaluated in the way presented in this paper.

The advantages of the the true Gaussian shapers are based on their better detection of closely overlapped short shaped pulses. The dead times of the true Gaussian and trapezoidal shapers are defined and calculated in Section \ref {sec:DeadResolvTime} with taking into account the output electronic noise and distortion of the shaped pulses. For large pulse widths, the calculated dead time coincides with the peaking plus flat top times of trapezoidal shapers. The concept of resolving time is introduced to specify the shortest time interval between two overlapped pulses when their mean inferred and input amplitudes differ less than the output shaper noise. The dead and resolving times, along with the amplitude measurement errors, are given in the aforementioned section as a function of the width of the shaped pulses. The algorithm of the pulse detection and measurement of their amplitudes is presented in Section \ref{sec:DeadTimeDetection}. The algorithm is applied for detection of true Gaussian and trapezoidal pulses at a high count rate in Section \ref{sec:Detection}. The results of the work are summarised in the Conclusion \ref{sec:Conclusion}.

\section{Dead and resolving times of true Gaussian and trapezoidal shapers}
\label{sec:DeadResolvTime}

The detection of closely overlapped photons and the maximal output count rate of the spectrometer are determined by its dead time $\tau_D$ \cite{Knoll2010,  AMPTEKSDD, AMPTEKGloss,  AMPTEKHigh}, i.e. the time interval after registration of a photon during which the system is not able to register any subsequent photons. The dead time of a trapezoidal shaper with an output pulse of undisturbed form and without pile-up rejection is the sum of the peaking and flat top times of the pulse, see (2) in \cite{AMPTEKSDD}. This sum equals the full width at half maximum (FWHM) of the trapezoidal pulse $T_s$. This definition becomes uncertain at small widths of shaped trapezoidal pulses $T_s\le\tau_r$ when its form is significantly disturbed and output noise is increased \cite{Kantor2019,Kantor2019Arx}. The peaking time is not directly applicable to the calculation of the dead time of the true Gaussian shapers either. Therefore, the dead time of a shaper is defined here with taking into account the form of the shaped pulses and the output noise of the shaper. It is valid for both the true Gaussian and standard shapers. The dead time is defined as the smallest time interval between two subsequent output pulses when: 

\begin{enumerate}[(1)]
	\item 	the sum of two single pulses has two maxima corresponding to the pulse peaks, separated by a valley; 
	\item 	the differences between the heights of both maxima and the valley are larger than the standard deviation of electronic noise at the shaper output. 
\end{enumerate}

\begin{figure}[H]
	\centering
	\includegraphics[width=\linewidth]{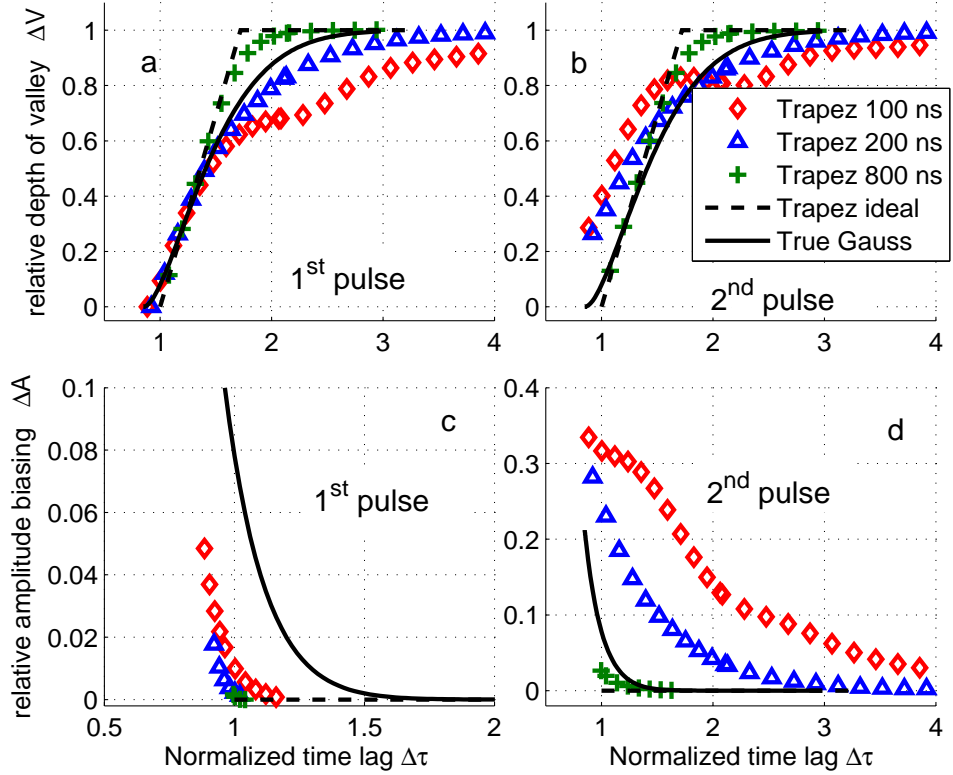}
	\caption{Relative depth of the valley between two overlapped pulses and biasing their amplitudes. Relative depth in regards with the amplitude of the first (a) and second (b) pulses, relative amplitude biasing of the first (c) and second (d) pulses}
	\label{fig:OverlappedPulses}
\end{figure}

The normalized relative difference $\Delta V=1-B/A_{1,2}$ between the maxima $A_{1,2}$ and the valley bottom $B$ are plotted in sections $a$ and $b$ of \figref{fig:OverlappedPulses} correspondingly for the first and second peaks of the overlapped shaped pulses. The pulses are converted from the impulse response of the AXAS-D system on an impact photon. The differences are plotted versus the time lag $T$ between the overlapped equal pulses normalized to the width of pulses at FWHM $\Delta \tau=T/T_s$.  

Black solid and dashed curves refer the differences $\Delta V$ calculated for a pair of equal pulses of the true Gaussian and ideal trapezoidal Gaussian-like forms. The latter form was introduced in \cite{Kantor2018}. This form is provided by a trapezoidal shaper \cite{jordanov1994digital, jordanov1994digital1} which time parameters are set so that to equate the widths and slopes of the trapezoidal pulse to those of the Gaussian pulse measured at the half maximum. The shaped pulse tends to the ideal trapezoidal form when its width is much greater than the rise time $\tau_r =203$ ns of the AXAS-D pulse \cite{Kantor2019}. 

At low noise, the dead times of the ideal trapezoidal  and Gaussian shapers equal $T_s$ and $\approx0.85T_s$ correspondingly as one can see in \figref{fig:OverlappedPulses} $a$ and $b$. The dead time of actual short trapezoidal pulses is close to that of the Gaussian pulses of the same width, see colour marks in the figure.  

The width of the trapezoidal pulses is restricted by $\approx$100 ns at FWHM because of strong distortion of its form \cite{Kantor2018}. This distortion results in biasing the maxima of overlapped trapezoidal pulses as shown in section $c$ and $d$ of \figref{fig:OverlappedPulses}. Here, the relative biasing $\Delta A=A_{1,2}/A_0-1$ between the detected maxima and the amplitude of the single pulse $A_0$ are plotted versus the normalized time lag. 

The ideal trapezoidal shaper provides biasing-free amplitudes of the overlapped pulses when the time lag is greater than the pulse width, $\Delta\tau>1$, as shown in dashed lines in plots $c$ and $d$ of \figref{fig:OverlappedPulses}. The overshoot tail of the first trapezoidal pulse produces a significant biasing amplitude of the second pulse in the pair at time lags much greater than the pulse width, $\Delta \tau >>1$. The biasing amplitudes of overlapped Gaussian pulses approach an 1\% level at the relative time lag $\Delta \tau =1.3$. 

Thus, accurate amplitude measurements of both trapezoidal and Gaussian pulses require a time lag between the pair of pulses somewhat larger than the dead time. We hereby define the resolving time $\tau_R$ of a pulse as the smallest time intervals to its preceding and successive neighbors when  
\begin{enumerate}[(3)]
	\item the difference between the inferred and actual amplitudes of the pulse is less than twice the standard deviation of electronic noise at the shaper output. 
\end{enumerate}

The preceding and successive resolving times of the true Gaussian and ideal trapezoidal pulses are equal because of their symmetrical form. The resolving time of short asymmetrical trapezoidal pulses splits in two branches - the preceding time is significantly larger than the successive time.

The dead and resolving times of true Gaussian and Gaussian-like trapezoidal shapers are estimated from the dependences $\Delta V$ and $\Delta A$ on the time lag calculated at a number of pulse widths while taking into account the output noise of the shapers. The numerical model considers the impulse response of the KETEK system on two subsequent photons of similar energies 5895 eV and a time lag $T$. The detector signals are digitized at a sampling frequency of 50~MHz and converted by true Gaussian or trapezoidal shapers to shorter pulses.

The statistical errors of the amplitude measurements of a single pulse are determined by the charge statistics of pair creation in the detector \cite{lechner1996pair,Schlosser2010} and the  electronic noise. The input electronic noise of the shapers is the normal white noise smoothed by a third-order low-pass filter to fit the experimental noise spectrum of the KETEK system shown in Fig. 1b in \cite{Kantor2018}. The standard deviation of the smoothed electronic noises was 55~eV. 

\begin{figure}[H]
	\centering
	\includegraphics[width=\linewidth]{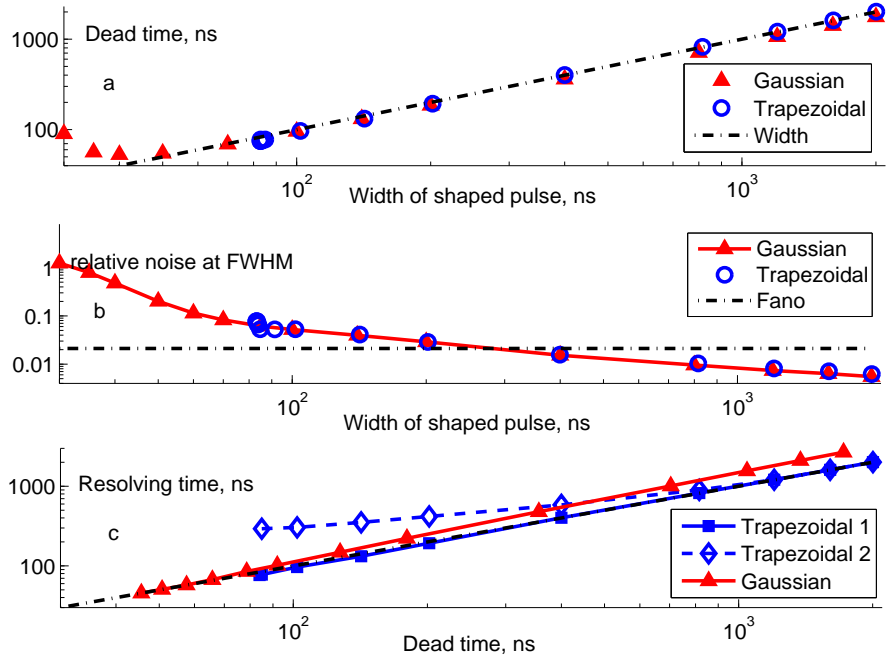}
	\caption{Dead times (a), relative electronic noise (b) and resolving times (c) of the true Gaussian and Gaussian-like trapezoidal shapers. The widths of shaped pulses are given in terms of FWHM.}
	\label{fig:DeadResolvTime}
\end{figure}

The dead times of the shapers, shown in \figref{fig:DeadResolvTime}a, were nearly equal to the widths of the shaped pulses at FWHM shown in a black dashed-dotted line  in the plot. The minimal dead time of the trapezoidal shaper is about a half of the rise time $\tau_r =200$ ns of the detector impulse response. The minimal dead time of the true Gaussian shaper is determined by the input signal noise. For the KETEK output pulse they are 90 ns and 40 ns correspondingly.  

The measured standard deviation of the electronic noise of the KETEK system amounted to 53~eV. The electronic noises of the detection system were modelled as white noise smoothed by a third-order low-pass filter with a cut-off frequency of 8~MHz. The spectrum of the modelled noises corresponded to that shown in Fig.1b in \cite{Kantor2018}. The output electronic noise of the shapers along with the Fano noise normalized to the pulse amplitude 5895~eV are shown in \figref{fig:DeadResolvTime}b. The Fano noise of peak amplitudes caused by the Fano statistics of electron hole pairs created by photon impacts in the detector area was estimated to be 130~eV at FWHM   \cite{lechner1996pair, Schlosser2010}. It corresponds well to the specifications the KETEK system with SDD of standard class \cite{KETEKVITUS}. The energy resolution of the shaped pulses is determined by the quadratic sum of the Fano statistics in the detector layer and the output electronic noise \cite{lechner1996pair}.

The resolving times of the true Gaussian and trapezoidal shapers are presented in \figref{fig:DeadResolvTime}c. The resolving time of the trapezoidal shaper splits  in two branches at small widths as shown in the plot. The solid curve with square marks labeled as 'Trapezoidal 1' relates to the successive resolving time, i.e. the least time interval to the second pulse of the pair which ensures the correct inferred amplitude of the first detected pulse. This amplitude is less affected by the second pulse and therefore the successive resolving time remains close to the dead time of the pair. The successive resolving time of the trapezoidal shaper is very close to its dead time. The second time relates to the amplitude resolution of the second pulse of the pair. Its amplitude can be strongly affected by a long tail of the first pulse. Therefore, the preceding resolving time remains large even at small dead times, see a blue dash curve with opened diamonds in the plot.

Thus, the trapezoidal shapers suit well for shaping to longer pulses and operating at moderate count rates, whereas they significantly loose their resolution at high count rates. Gaussian shapers are advantageous at a high count rate when operating  with short shaped pulses. True Gaussian shaping reduces the least resolving time from 300~ns down to 50~ns. This allows a significant increase of the output count rate of the KETEK spectrometer. Smaller dead and resolving times of the true Gaussian shaper give an advantage of better detection of close pulses and better distinguishing the pile-up signals. Detection of true Gaussian and trapezoidal pulses was tested in a wide range of input count rates with a numerical model described in the next section. 

\section{Detection of true Gaussian and trapezoidal pulses}
\label{sec:DeadTimeDetection}

The impulse response of the KETEK spectrometer was measured with the use of a radiation source $^{55}Fe$. The maximal input count rate provided by the source is $\approx 10^4$ 1/s, therefore detection at higher rates up to $10^7$ 1/s was simulated numerically using the characteristics of the KETEK system measured at lower count rates.  

The radiation source was simulated to emit photons of 5895~eV energy, distributed randomly and uniformly in a given time interval. The inverse mean time interval between subsequent photons which impact the SDD sensor is the input count rate of the shapers. The output count rate of a detection system is defined as the inverse mean time interval between subsequent detected or resolved impulse responses. 

The simulated impulse responses at the output of the KETEK system were synchronised with the impact photons. The response amplitudes are normally distributed around the mean photon energy with the standard deviation of the Fano noises 55 eV \cite{Schlosser2010}. The sum of the simulated impulse responses and noises were converted by the true Gaussian and trapezoidal shapers to a sequence of shorter pulses in the way described above. 

A threshold level is introduced to separate the noise and actual signal. It is found from the histograms of the noise measured prior the first impact photon. The threshold is defined at the level when the high amplitude wing of the noise histogram reduces to the 50\% of the signal+noise histogram. 

\begin{figure}[H]
	\centering
	\includegraphics[width=\linewidth]{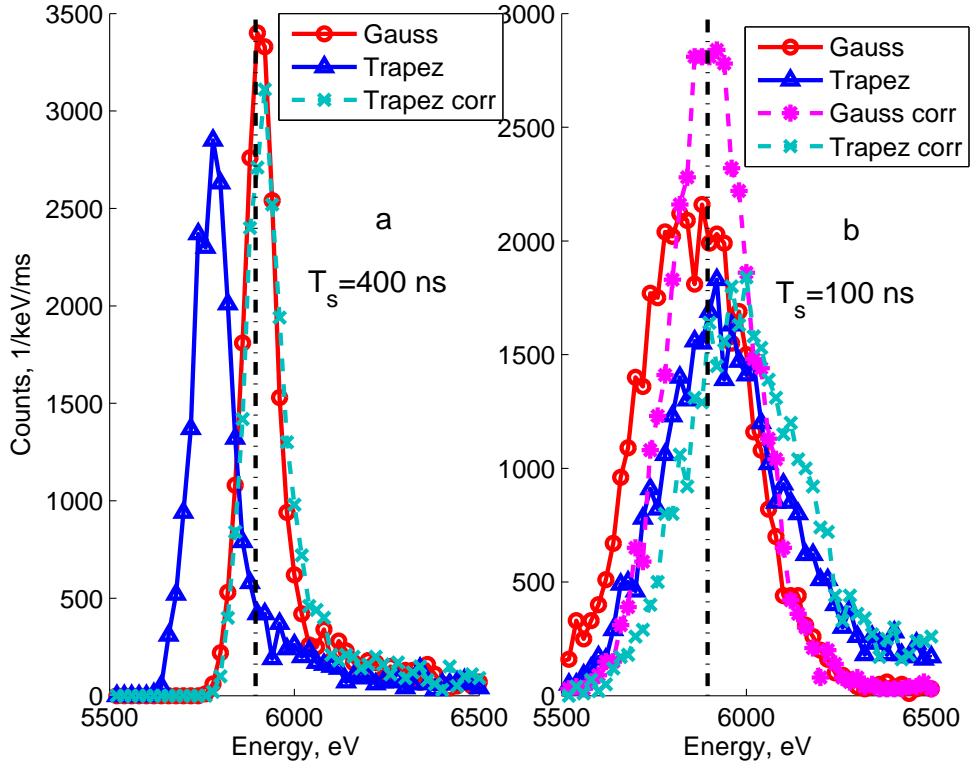}
	\caption{Spectra of resolved and corrected pulse amplitudes of the true Gaussian and trapezoidal shapers at the input count rate of $10^6 1/s$} 
	\label{fig:Spectra}
\end{figure}

The shaped signals were analysed using a numerical code which returns the times and amplitudes of the peaks detected above the threshold level. The spectra of the detected pulses with amplitudes near the photon energy are shown in \figref{fig:Spectra}. There are 5000 AXAS-D pulses of a 5895 eV amplitude generated against the electronic noise of AXAS-D at the mean count rate of $10^6$ 1/s. The pulses were digitized at a 50 MHz sampling frequency. The Fano noise of the detector was put to zero for comparison of the amplitude spectra of Gaussian and trapezoidal shapers.

Red and blue solid curves with open circles and triangles correspond to the measured amplitudes of the Gaussian and trapezoidal pulses of 400 ns and 100 ns widths. The photon energy is marked in black dashed-dotted lines in the figure. The spectrum of Gaussian pulses of 400 ns width is well centered to the photon energy. Its skewness to the high energy side is accounted for overlapping the shaped pulses.  

\begin{figure}[h]
	\centering
	\includegraphics[width=\linewidth]{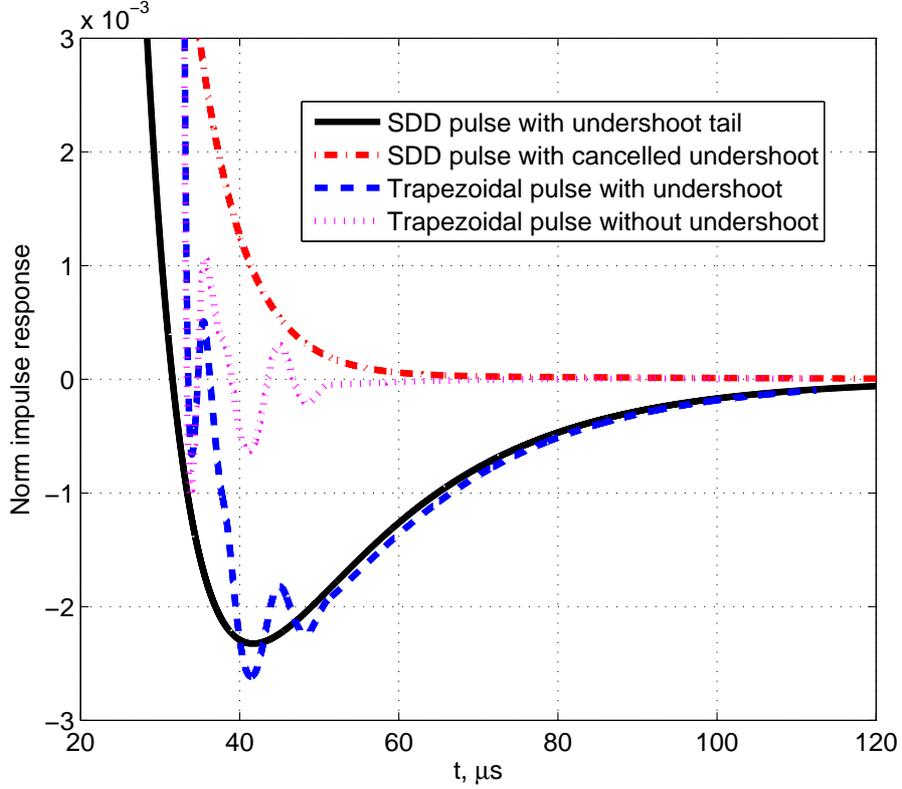}
	\caption{Tails of the KETEK and shaped pulses with and without digital pole-zero cancellation}
	\label{fig:UndershootCancellation}
\end{figure}

The spectrum of trapezoidal pulses is significantly biased in the opposite direction. This large negative biasing is caused by the accumulation of long undershoot tails of many pulses. These undershoots come from a 0.2 \% undershoot of the $S_R$ pulse shown in Fig. 1a in \cite{Kantor2018}. The undershoot of this pulse is not perfectly cancelled with the pole-zero circuit of the preamplifier \cite{Knoll2010}. The complete cancellation was done with a digital reversed pole-zero circuit which was found to significantly reduce the tail of the KETEK spectrometer impulse response, as shown in \figref{fig:UndershootCancellation}. The measured undershoot tail is shown in a solid black curve. The output tail after its digital cancellation is plotted in a dashed-dotted red curve. The tails of the trapezoidal pulses converted from the measured and filtered $S_R$ pulses are plotted in dashed blue and dotted magenta curves, respectively. The cancellation filter allows for a reduction of biasing in the trapezoidal amplitudes down to the biasing of Gaussian pulses and does not change other characteristics of the trapezoidal shapers. The amplitude spectrum of the corrected trapezoidal pulses is plotted in \figref{fig:Spectra}a in a cyan dashed curve with cross marks. The spectrum of the corrected Gaussian amplitudes is very similar to the measured one and therefore is not plotted in the figure.  Hereafter the output KETEK pulse with undershoot cancellation is used for analysis of the shapers.

The spectra of measured amplitudes of short 100 ns shaped pulses are shown in \figref{fig:Spectra}b. One can see a negative biasing of the Gaussian spectrum plotted in a solid red curve with open circles. It is caused by a small digitizing period $\tau_s$ of the input signals in respect to the width of the shaped pulses $T_s$. The measured amplitudes of shaped pulses are corrected when their width is less then $\approx\tau_s\sqrt{A_0/\sigma_{noise}}$  at FWHM, where $A_0$ is the mean peak amplitude and $\sigma_{noise}$ is the rms of the noise.     

The amplitude of a single true Gaussian pulse was corrected in two steps. First, the time of the peak $t_0$ was found from the parabolic approximation of  signals $S_i$ around the detected peak maxima. The amplitude of the Gaussian pulse $A_{Gauss}$ was found from the peak time $t_0$ and the known pulse width $T_s$ with the use of the least square root regression of the top part of single Gaussian pulses:

\begin{align}
&  A_{Gauss}=\frac{\sum S_iexp(-K((t_i-t_0)/T_s)^2)}{\sum exp(-2K((t_i-t_0)/T_s)^2)}  
&  K=4\log 2    
\end{align}
 
The top parts of overlapped true Gaussian pulses do not have the Gaussian form, therefore they were approximated by polynomials from the 3rd to 6th power for amplitude corrections. The same approximation was applied for correction of amplitudes of trapezoidal pulses. The corrected spectra are plotted in the figure in dashed magenta and cyan curves with star and cross marks for the Gaussian and trapezoidal shapers. The correction removes the bias of the Gaussian pulse amplitudes and improves their spectral resolution by 30\% . The amplitude correction reveals biasing amplitudes of trapezoidal pulses in the high energy side caused by the overshoot of the pulse tails. 

The algorithm was applied for simulation of the output count rate,  energy resolution and amplitude biasing of the shapers at different input count rates. The results are presented in the next section. 

\section{Detection of shaped pulses at a high count rate}
\label{sec:Detection}

The detector output signals were shaped in three channels for detection of true Gaussian and trapezoidal pulses. A true Gaussian shaper with output pulses of 50 ns width was served as a fast channel for pile-up rejection of peaks detected in other two slow channels as described in \cite{AMPTEKHigh,Goulding1983}. Conditions (1) and (2) cannot be directly used for selection of pulses in the fast channel because times of impact photons are unknown. The threshold limit of peak amplitudes and the least pulse width of the Gaussian peaks were employed for the selection of the detected peaks instead. The count rate of the detected Gaussian pulses in the fast channel $C_{out}$ fits well to the theoretical value  $C_{out}=C_{in}\cdot exp(-C_{in}T_s)$ \cite{Knoll2010, AMPTEKSDD}, where $C_{in}$ is the input count rate and $T_s$=50 ns. The efficiency of the pulse detection in the fast channel achieves 80\% at $C_{in}=5\cdot10^6$  1/s.

Two slow channels equipped with true Gaussian and trapezoidal shapers of the same widths of the output pulses are used for detection of shaped pulses of larger widths. The pulses in the slow channel were detected using the threshold limit, the least pulse width and their synchronization with the peaks in the fast channel. The resolved pulses were selected with the use of condition (3), i.e.  the least time intervals to the neighbor peaks in the fast channel. The peaks separated less than the resolving times were rejected from the counted set. The preceding and successive resolving times were applied for the pile-up rejection of trapezoidal pulses. The remained peaks are referred hereafter to as the resolved peaks. The output count rate, energy resolution and biasing of the resolved peaks were calculated for 1000 input pulses. Therefore, the simulated data have some spread around their mean values.   

\begin{figure}[H]
	\centering
	\includegraphics[width=\linewidth]{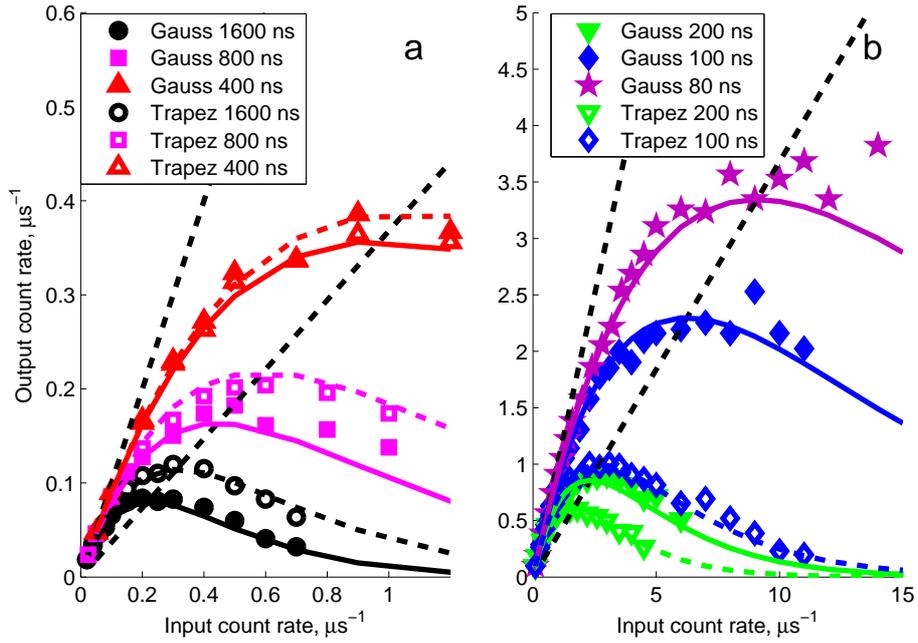}
	\caption{Output count rate with true Gaussian and Gaussian-like trapezoidal shapers for large (a) and small (b) shaping widths.}
	\label{fig:CountRates}
\end{figure}

The count rates of the resolved peaks of the true Gaussian and trapezoidal shapers are shown in two plots in \figref{fig:CountRates} for large and small shaping widths versus the input count rate of the simulated detector pulses. The solid and open marks relate to the true Gaussian and trapezoidal shapers correspondingly. 

The first plot in the figure presents simulations of the count rates for shaping widths larger than the twice rise time of the $S_R$ pulse $\approx$400 ns. The resolving time of true Gaussian pulses at those widths is larger than the mean of the preceding and successive resolving times of trapezoidal pulses. The resolving time of true Gaussian pulses in the second plot is less than the mean of the preceding and successive resolving times of trapezoidal pulses. 

The well known theoretical dependence between the output and input rates for symmetrical shaped pulses is $C_{out}=C_{in}\cdot exp(-2C_{in}\tau_R)$ \cite{Knoll2010, AMPTEKSDD}. This dependence generalised to asymmetrical pulses by substituting $2\tau_R=\tau_{Rp}+\tau_{Rs}$, where $\tau_{Rp}$ and $\tau_{Rs}$ are the preceding and successive resolving times. The formula is derived in the assumption that the pulses detected in the fast channel and used for pile-up rejection have a uniform random time distribution. It is true while the resolving time in the slow channel is much greater than the dead time in the fast channel. Otherwise, the formula must be corrected for the finite dead time $\tau_{Df}$ in the fast channel \cite{Abbene2015}:

\begin{align}
C_{out}=C_{in}\cdot exp(-C_{in}(\tau_{Rp}+\tau_{Rs}-\tau_{Df}))  
\end{align}        

These theoretical rates are plotted in the figure in curves with colours corresponding to the marks of the simulated data. The solid curves regard with the Gaussian and dashed curves refer to trapezoidal shapers. The output rates $C_{out}$ that equals the input rates $C_{in}$ are plotted in the upper black dashed straight line. The theoretical maximum of the output count rates $C_{in}/e$ is plotted in the lower black dashed straight lines. 

Formula (2) does not take into account reset pulses periodically applied to discharge the capacitor which collects the electron charge created by the detected photons. When the accumulated photon energy achieves a few MeV, the capacitor is shortcircuited photons are not counted a few microseconds. Hundreds photons of 6 keV energy can be detected between two reset pulses. At the highest photon rates $10^7$~1/s, the collection time of these photons is $\approx 100 \mu s$. So, the reset pulses can increase the dead time by $\approx 5\%$ at the highest count rates. The output count rate is calculated in the paper for photons detected between two resets to eliminate the effect of the reset pulses on the photon count rate.  

The simulated count rates well fit to the theoretical predictions in a wide range of the pulse widths and input count rates. At large widths of trapezoidal pulses, $T_s>400 ns$, they accord also with the characteristics of the AXAS-D spectrometers \cite{KETEKAXAS} with the impulse responses of corresponding peaking times. The count rates of Gaussian pulses are less in this range because of their larger resolving time, see \ref{fig:DeadResolvTime}c. At small shaping widths, the true Gaussian shapers are superior and provide a higher count rate. The count rate of Gaussian pulses is additionally boosted to three times the rate of the trapezoidal shapers when the pulse widths in the fast and slow channels are getting closer, see (2). 

\begin{figure}[!h]
	\centering
	\includegraphics[width=\linewidth]{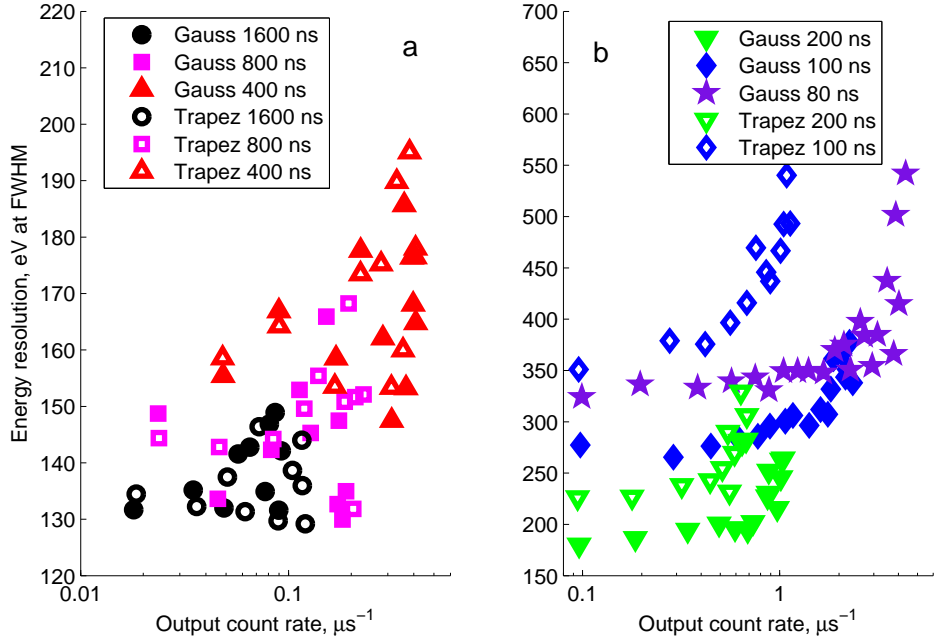}
	\caption{Amplitude resolution at FWHM of pulses resolved with true Gaussian and  Gaussian-like trapezoidal shapers for large (a) and small (b) shaping widths.}
	\label{fig:Resolution}
\end{figure}

The maximal output count rate of 100 ns Gaussian pulses is $2.3\cdot10^6$~counts per second, which is a factor of 2.3 higher than that calculated for the trapezoidal pulses of the same width and 7 times higher than the maximal output rate of the AXAS-D spectrometer with modern Vitus H7 detector of Cube class\cite{KETEKVITUSCube}. The highest output rate of resolved Gaussian pulses $3.5\cdot10^6$~counts per second is achieved at the pulse width 80~ns. Higher rates are restricted by the output noise of the shaper. 

The output count rates of short Gaussian pulses accord well with the rates presented for a fast and low noise AMPTEK SDD spectrometer \cite{AMPTEKHigh} designed for operation at output count rates up to $10^6$~1/s. The rise time of the step-like detector signals in the AMPTEK spectrometer is about one tenth of that in the AXAS-D spectrometer and therefore short shaped pulses could be provided by conventional trapezoidal shapers. The slower AXAS-D spectrometer with the true Gaussian shaper is capable to exceed the maximal count rate of the fast AMPTEK spectrometer by a factor of 2. One could expect a significant rise of the count rate of the AMPTEK spectrometer equipped with the true Gaussian shaper.  
  
The amplitude resolution of the resolved pulses in terms of FWHM of the main spectral peak is shown in \figref{fig:Resolution} versus the output count rate of the shapers. The data are marked in the same way as in \figref{fig:CountRates}. The resolutions of the Gaussian and trapezoidal pulses at low count rates are in the range of the technical specifications of the tested KETEK spectrometer providing long shaped pulses $T_s>400$~ns \cite{KETEKAXAS,KETEKVITUS}. It drops when the output count rate approaches their maxima. The amplitude resolution of shorter Gaussian pulses $T_s<400$~ns  get better than that of  trapezoidal pulses of the same width. The found energy resolution of true Gaussian pulses is $\approx 25\%$ better than the resolution provided by the low noise AMPEK spectrometer with trapezoidal shapers with similar peaking times \cite{AMPTEKHigh}.

\begin{figure}[H]
	\centering
	\includegraphics[width=\linewidth]{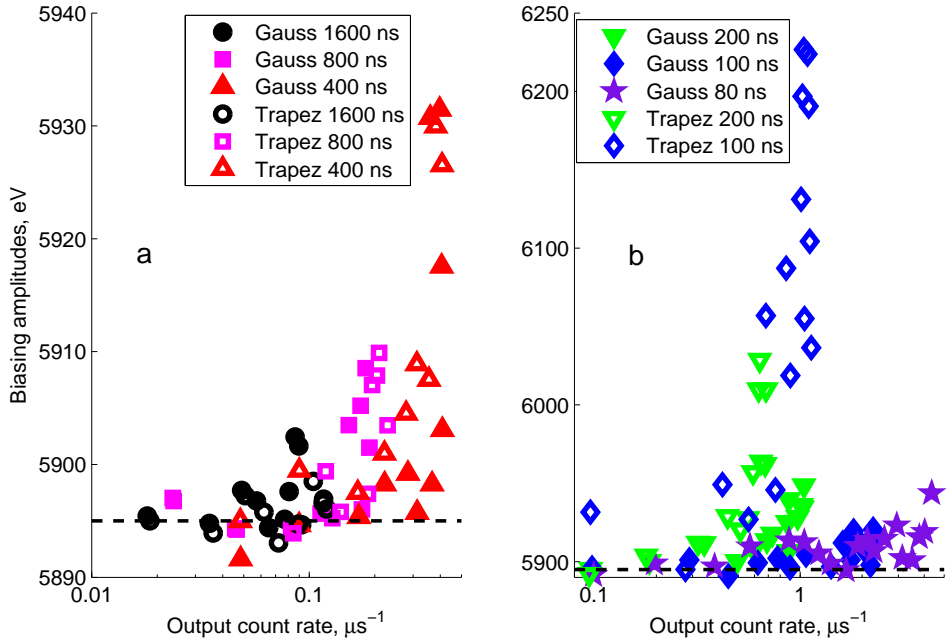}
	\caption{Biasing pulse amplitudes of true Gaussian and  Gaussian-like trapezoidal shapers for large (a) and small (b) shaping widths.}
	\label{fig:Bias}
\end{figure}

Another important advantage of the true Gaussian shapers is biasing-free measurements of pulse amplitudes, see \figref{fig:Bias}. Both shapers provide similar biasing $<50$ eV for pulse widths larger than 200 ns. Biasing shorter Gaussian pulses is kept in the same range up to the output count rates $\approx 4\cdot10^6$ 1/s. Overlaps of short distorted trapezoidal pulses results in their high biasing amplitudes at the output count rates larger than $\approx 7\cdot 10^5$ 1/s. 

Thus, the true Gaussian shaper increases the maximal count rate of AXAS-D spectrometer by a factor of 7. The maximal count rate of the AXAS-D spectrometer equipped with a true Gaussian shaper is a twice the rate of the fast AMPTEK spectrometer. The amplitude resolution of true Gaussian pulses is $\approx$25\% better than the resolution of trapezoidal pulses at short shaping widths.

\section{Conclusion}
\label{sec:Conclusion}

A new digital true Gaussian shaper developed for pulse counting at high rates has been analysed in terms of the dead and resolving times, output count rate and pulse amplitude energy resolution. The analysis employed numerical models based on calibrations and measured characteristics of a soft x-ray spectrometer AXAS-D with a silicon drift detector H7 VITUS of standard class developed by KETEK GmbH. The proposed true Gaussian shaper provides the following advantages in regards with standard trapezoidal shapers:

\begin{enumerate}[(1)]
	\item 	symmetrical output pulses whose width can be shorter than the rise time of the detector impulse response; 
	\item 	better resolution of the overlapped detector pulses at very small time lags; 
	\item 	several time higher count rates at small pulse widths; 
	\item   better amplitude resolution at high count rates; 
	\item   biasing-free measurements of pulse amplitudes at high count rates.   
\end{enumerate}

\section*{Acknowledgments}
This work was supported by Ioffe Institute in Saint-Petersburg, Russia.
 
\bibliography{Kantor-GaussDetect-Arxiv}{} 

\begin{thebibliography}{10}
\expandafter\ifx\csname url\endcsname\relax
  \def\url#1{\texttt{#1}}\fi
\expandafter\ifx\csname urlprefix\endcsname\relax\def\urlprefix{URL }\fi
\expandafter\ifx\csname href\endcsname\relax
  \def\href#1#2{#2} \def\path#1{#1}\fi

\bibitem{Knoll2010}
G.~F. Knoll, Radiation Detection and Measurement, 4th Edition, John Wiley and
  Sons, Inc., Hoboken, NJ, USA, 2010.

\bibitem{jordanov1994digital}
V.~T. Jordanov, G.~F. Knoll, Digital synthesis of pulse shapes in real time for
  high resolution radiation spectroscopy, Nuclear Instruments and Methods in
  Physics Research Section A: Accelerators, Spectrometers, Detectors and
  Associated Equipment 345~(2) (1994) 337--345.

\bibitem{jordanov1994digital1}
V.~T. Jordanov, G.~F. Knoll, A.~C. Huber, J.~A. Pantazis, Digital techniques
  for real-time pulse shaping in radiation measurements, Nuclear Instruments
  and Methods in Physics Research A 353~(1) (1994) 261--264.

\bibitem{Kantor2019}
M.~Y. Kantor, A.~V. Sidorov, Shaping pulses of radiation detectors into a true
  gaussian form, Journal of Instrumentation 14~(1) (2019) 1004.

\bibitem{Kantor2018}
M.~Y. Kantor, A.~V. Sidorov, True gaussian shaping for high count rate
  measurements of pulse amplitudes (2018).
\newblock \href {http://arxiv.org/abs/arXiv:1809.02211}
  {\path{arXiv:arXiv:1809.02211}}.

\bibitem{Kantor2019Arx}
M.~Y. Kantor, A.~V. Sidorov, Detection of true gaussian shaped pulses at high
  count rates (2019).
\newblock \href {http://arxiv.org/abs/arXiv:1909.10875}
  {\path{arXiv:arXiv:1909.10875}}.

\bibitem{KETEKAXAS}
\href{http://www.ketek.net/products/axas/axas-d}{{AXAS-D Analytical X-ray
  Acquisition System - KETEK GmbH}}.
\newline\urlprefix\url{http://www.ketek.net/products/axas/axas-d}

\bibitem{KETEKVITUS}
\href{https://www.ketek.net/wp-content/uploads/2017/10/KETEK_Manual_VITUS_SDD_REV6_2017-10.pdf}{{VITUS
  Silicon Drift Detector. User's manual - KETEK GmbH}}.
\newline\urlprefix\url{https://www.ketek.net/wp-content/uploads/2017/10/KETEK_Manual_VITUS_SDD_REV6_2017-10.pdf}

\bibitem{KETEKVITUSCube}
\href{https://www.ketek.net/sdd/vitus-sdd-modules/vitus-h7/}{{VITUS Silicon
  Drift Detector of Cube class}}.
\newline\urlprefix\url{https://www.ketek.net/sdd/vitus-sdd-modules/vitus-h7/}

\bibitem{AMPTEKSDD}
\href{https://www.amptek.com/-/media/ametekamptek/documents/resources/ansdd1.pdf}{{Amptek
  Silicon Drift Diode (SDD) at High Count Rates}}.
\newline\urlprefix\url{https://www.amptek.com/-/media/ametekamptek/documents/resources/ansdd1.pdf}

\bibitem{AMPTEKGloss}
\href{https://www.amptek.com/-/media/ametekamptek/documents/resources/glossary.pdf}{{AMPTEK
  General tutoirials. Glossary}}.
\newline\urlprefix\url{https://www.amptek.com/-/media/ametekamptek/documents/resources/glossary.pdf}

\bibitem{AMPTEKHigh}
\href{https://www.amptek.com/-/media/ametekamptek/documents/resources/andpp2.pdf}{{Amptek
  Operating the DP5 at High Count Rates}}.
\newline\urlprefix\url{https://www.amptek.com/-/media/ametekamptek/documents/resources/andpp2.pdf}

\bibitem{lechner1996pair}
P.~Lechner, R.~Hartmann, H.~Soltau, L.~Str{\"u}der, Pair creation energy and
  fano factor of silicon in the energy range of soft x-rays, Nuclear
  Instruments and Methods in Physics Research A 377~(2-3) (1996) 206--208.

\bibitem{Schlosser2010}
D.~Schlosser, P.~Lechner, et.al, Expanding the detection efficiency of silicon
  drift detectors, Nuclear Instruments and Methods in Physics Research A 624
  (2010) 270--276.

\bibitem{Goulding1983}
F.~S. Goulding, D.~A. Landis, N.~W. Madden,
  \href{https://ieeexplore.ieee.org/stamp/stamp.jsp?arnumber=4332275}{Design
  philosophy for high-resolution rate and throughput spectroscopy systems},
  IEEE Transactions on Nuclear Science 30~(1) (1983) 301--310.
\newblock \href {https://doi.org/10.1109/TNS.1983.4332275}
  {\path{doi:10.1109/TNS.1983.4332275}}.
\newline\urlprefix\url{https://ieeexplore.ieee.org/stamp/stamp.jsp?arnumber=4332275}

\bibitem{Abbene2015}
L.~Abbene, G.~Gerardi, High-rate dead-time corrections in a general purpose
  digital pulse processing system, Journal of Synchrotron Radiation 22~(5)
  (2015) 1190--1201.

\end{thebibliography}
\end{document}